\documentclass[preprint, amsmath, amssymb, preprintnumbers, showpacs, showkeys, aps, prb]{revtex4-1}

\usepackage{graphicx}
\usepackage{float}
\usepackage{color}
\usepackage{natbib}
\begin{document}

\title{High thermoelectric figure of merit in $p$-type Half-Heuslers by intrinsic phase separation}

\author{Elisabeth Rausch}
\affiliation{Institut f\"ur Anorganische und Analytische Chemie, Johannes Gutenberg - Universit\"at, 55099 Mainz, Germany}
\affiliation{Max Planck Institute for Chemical Physics of Solids, D-01187 Dresden, Germany}

\author{Siham Ouardi},
        \affiliation{Max Planck Institute for Chemical Physics of Solids, D-01187 Dresden, Germany}

\author{Ulrich Burkhardt}
       \affiliation{Max Planck Institute for Chemical Physics of Solids, D-01187 Dresden, Germany}
               
\author{Claudia Felser}
       \affiliation{Max Planck Institute for Chemical Physics of Solids, D-01187 Dresden, Germany}
       
\author{Jana Marie Stahlhofen}
         \affiliation{Institut f\"ur Anorganische und Analytische Chemie, Johannes Gutenberg - Universit\"at, 55099 Mainz, Germany}
\author{Benjamin Balke}
       \email{balke@uni-mainz.de}
       \affiliation{Institut f\"ur Anorganische und Analytische Chemie, Johannes Gutenberg - Universit\"at, 55099 Mainz, Germany}
       

\date{\today}

\begin{abstract}
Improvements in the thermoelectric properties of Half-Heusler materials have been achieved by means of a micrometer-scale 
phase separation that increases the phonon scattering and reduces the lattice thermal conductivity. 
A detailed study of the $p$-type Half-Heusler compounds Ti$_{1-x}$Hf$_{x}$CoSb$_{0.85}$Sn$_{0.15}$ 
using high-resolution synchrotron powder X-ray diffraction and element mapping electron microscopy 
evidences the outstanding thermoelectric properties of this system.
A combination of intrinsic phase separation and adjustment of the carrier concentration via Sn substitution 
is used to realize a record thermoelectric figure of merit for $p$-type Half-Heusler compounds of $ZT$ =~1.15 at 710~$^{\circ}$C in  Ti$_{0.25}$Hf$_{0.75}$CoSb$_{0.85}$Sn$_{0.15}$.
The phase separation approach can form a significant alternative to nanostructuring processing time, 
energy consumption and increasing the thermoelectric efficiency.

\end{abstract}

\maketitle

\section{Introduction}
Intermetallic Half-Heusler compounds with the structural formula $XYZ$ and with the cubic structure 
$F \overline{4}3m$ have recently gained attention as promising materials for moderate-temperature 
thermoelectric (TE) applications such as industrial and automotive waste heat recovery.\citep{BD14}
An industrial upscaling of the synthesis of both $p$-type and $n$-type materials has been realized, 
and such thermoelectric modules have been tested for their long-term stability and reproducibility.\citep{BBZ14} 
However, their performance still needs to be improved by adoption of the optimal material composition
to reach the cost efficiency of industrial applications.\\
The efficiency of TE materials is characterized by its figure of merit $ZT = S^2\sigma T / \kappa $, 
where $S$ denotes the Seebeck coefficient, $\sigma$ the electrical conductivity, $T$ the absolute Temperature
and $\kappa$ the thermal conductivity, which is composed of the 
lattice component $\kappa_{lat}$ and the electronic component $\kappa_{el}$.
Since the Seebeck coefficient $S$, the electrical conductivity $\sigma$ and the electronic thermal 
conductivity $\kappa_{el}$ are interrelated through the carrier concentration $n$, these parameters 
cannot be manipulated separately. High power factors $S^2 \sigma$ are generally achieved in heavily 
doped semiconductors with $n$ in the range from 10$^{19}$~cm$^{-3}$ to 10$^{21}$~cm$^{-3}$ depending 
on the material system.\citep{ST08}\\
A great advantage of Half-Heusler compounds is the possibility to substitute each of the three 
occupied fcc sublattices individually.
For example, it is possible to tune the number of charge carriers by substitution of the $Z$-position element
by another main-group element and simultaneously introduce disorder by substitution on the $X$- and 
$Y$-position elements resulting in mass fluctuations, which can decrease the thermal conductivity $\kappa$.\citep{GFP11}
In general, Half-Heusler compounds exhibit high power factors in the range of 2--6$\times$10$^{-3}$Wm$^{-1}$K$^{-1}$.\citep{GFP11,XWT12,CR13,BD14} 
Thus, the main obstacle to further improving their TE performance is their relatively high thermal conductivity. 
Recently, the concept of an intrinsic phase separation has become a focus of research. In particular, the 
$n$-type (Ti/Zr/Hf)NiSn system has been investigated in this regard\citep{KSS14,GZ14,GPS13,SB13,DMB13,KUM09} leading to several patent applications.\citep{Bosch,Toshiba,Shutoh,Graf} Very recently the stability of the submicrostructuring and thermoelectric properties under thermal cycling conditions was prooved.\citep{Krez} \\
In contrast, state-of-the-art $p$-type Half-Heusler materials rely on a nanostructuring approach involving ball 
milling followed by a rapid consolidation method which is a very time and energy consuming synthesis route.\citep{YLW12,YLC13} We have recently demonstrated that 
the concept of phase separation can also be applied to the $p$-type material system (Ti/Zr/Hf)CoSb$_{0.8}$Sn$_{0.2}$.\citep{RBO14}\\
Thus far, the best $p$-type Half-Heusler materials are based on a substitution of 20\% Sn on the Sb-position 
in the (Ti/Zr/Hf)CoSb material system.\citep{YJL11,YLW12,CSP08,HLM14} 
An additional adjustment of the carrier concentration further enhances the power 
factor. Optimum is a substitution level of 15\% Sn on the Sb-position.\citep{RB-sub}
This work combines both concepts in the investigation of the Ti$_{1-x}$Hf$_{x}$CoSb$_{0.85}$Sn$_{0.15}$ system. 
A fine-tuning of the TE performance via the adjustment of the Ti to Hf ratio is possible and the resulting TE properties 
correlate to the results of the structural investigations.

\section{Results and discussion}
\subsection {Microstructure investigations}
The investigations of the samples by X-ray powder diffraction (XPD), scanning electron microscopy (SEM), 
and energy-dispersive X-ray (EDX) spectroscopy revealed that samples containing Ti and Hf undergo 
an intrinsic phase separation into two Half-Heusler phases. The composition of the matrix (phase I) 
and the second Half-Heusler phase (phase II) as determined by EDX spectroscopy are indicated in Table~\ref{tbl:edx}.

\begin{figure}[ht]
\centering
   \includegraphics[width=15cm]{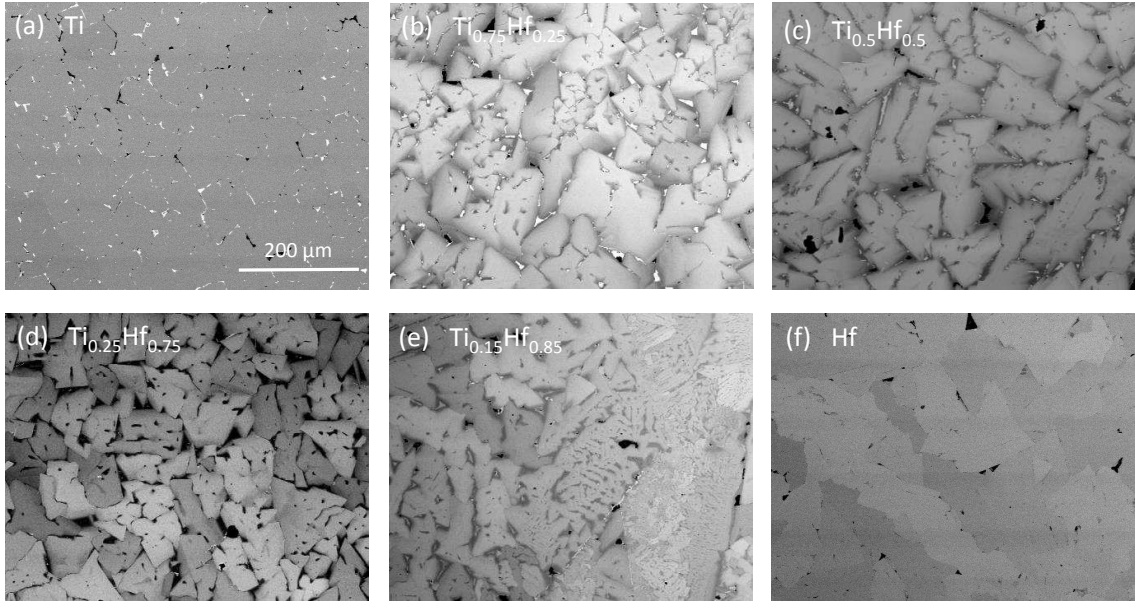}
     \caption{Backscattered electron images of Ti$_{1-x}$Hf$_{x}$CoSb$_{0.85}$Sn$_{0.15}$ with the indicated ratios of Ti to Hf. 
     					The matrix (mid-scale grey) is interlaced by a second Half-Heusler phase (dark regions), and the bright spots indicate Sn inclusions.}
  \label{fgr:sem}
\end{figure}

In general, we note that the matrix is rich in Hf and Sb whereas the second phase 
is rich in Ti and Sn when compared with the nominal composition. Moreover, this second phase has a slight excess of Co. 
We assume that the Co atoms occupy the vacant tetrahedral holes in the structure that is similar to the Heusler 
compound Co$_{2}$TiSn (L2$_{1}$ structure).\citep{BOG10}
 
Our assumption is supported by a recent report on the in-situ growth of Heusler quantum dots in 
the Half-Heusler Ti$_{0.5}$Hf$_{0.5}$CoSb$_{0.8}$Sn$_{0.2}$ matrix.\citep{SLM13}
In addition, very small amounts of the pure elements Ti, Sn or Hf or a 
binary Ti$_{y}$Sn$_{z}$ were detected. These results agree well with our previous investigations of the phase 
separation in the (Ti/Zr/Hf)CoSb$_{0.8}$Sn$_{0.2}$ system.\citep{RBO14}

\begin{figure}[ht] 
\centering
   \includegraphics[width=15cm]{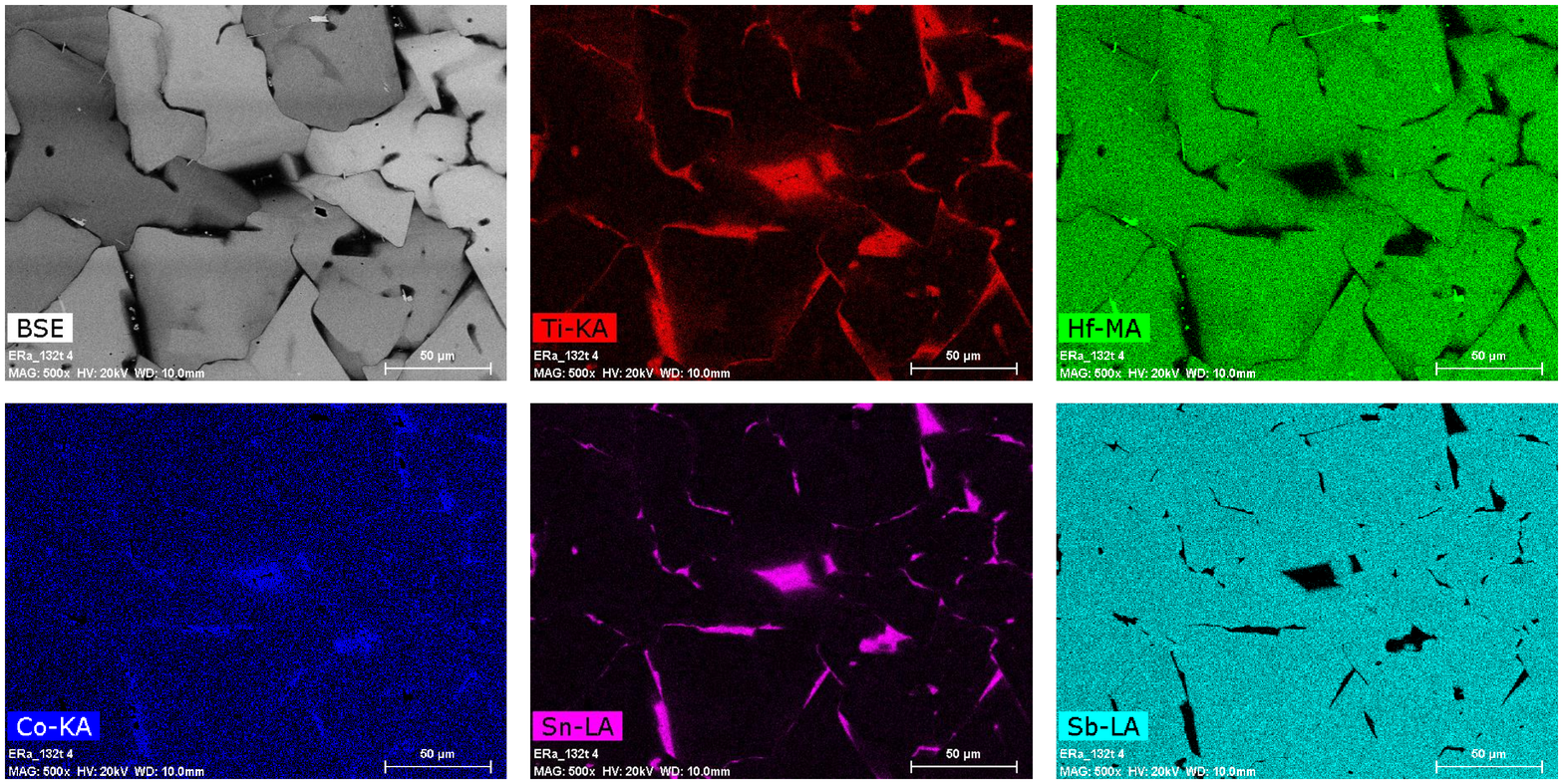}
   \caption{Element-specific EDX mapping of the sample Ti$_{0.25}$Hf$_{0.75}$CoSb$_{0.85}$Sn$_{0.15}$.}
  \label{fgr:mapping}
\end{figure}

The typical microstructure of the samples is displayed in the backscattered electron 
images in Figures~\ref{fgr:sem}(a)-(f). The amount and shape of the second Half-Heusler phase 
depends on the ratio of Ti to Hf.
Samples with 0.75 $\ge$ x $\ge$ 0.25 show a fine network of the second phase within the matrix. 
The most distinctive formation is observed for the composition with $x = 0.25$.
Upon decreasing the amount of Ti further ($x = 0.85$), the amount of the second Half-Heusler 
phase becomes too less to form an interconnecting network. 
The element-specific EDX mappings (see Figure~\ref{fgr:mapping} for the sample with x = 0.25) 
confirm that the second Half-Heusler phase is rich in Ti and Sn and exhibits 
a slight excess of Co. The images of the distributions of these three elements all 
appear similar, whereas the element-specific mappings for Hf and Sb
appear as "negatives" of the images of the other three elements. 


\begin{table}[ht]
\small
  \caption{\ Composition of the matrix (I) and second Half-Heusler phases (II) of the samples Ti$_{1-x}$Hf$_{x}$CoSb$_{0.85}$Sn$_{0.15}$ as determined by EDX spectroscopy. Experimental density $\rho$.
  }
  \label{tbl:edx}
\begin{tabular}{lclr}
Nominal composition  &    Phase   & EDX & $\rho$ / g$\times$cm$^{-3}$\\
\hline
TiCoSb$_{0.85}$Sn$_{0.15}$    & I     & Ti$_{1.07}$Co$_{0.98}$Sb$_{0.91}$Sn$_{0.05}$ & 7.312\\
      & II    & Ti$_{0.97}$Co$_{1.18}$Sb$_{0.28}$Sn$_{0.57}$& \\
Ti$_{0.75}$Hf$_{0.25}$CoSb$_{0.85}$Sn$_{0.15}$ & I     & Ti$_{0.79}$Hf$_{0.25}$Co$_{0.97}$Sb$_{0.94}$Sn$_{0.04}$ & 8.161 \\
      & II    & Ti$_{0.95}$Hf$_{0.09}$Co$_{1.06}$Sb$_{0.36}$Sn$_{0.55}$ &\\
Ti$_{0.5}$Hf$_{0.5}$CoSb$_{0.85}$Sn$_{0.15}$ & I     & Ti$_{0.44}$Hf$_{0.55}$Co$_{1.02}$Sb$_{0.94}$Sn$_{0.05}$ & 8.983 \\
      & II    & Ti$_{0.80}$Hf$_{0.21}$Co$_{1.07}$Sb$_{0.33}$Sn$_{0.59}$ &\\
Ti$_{0.25}$Hf$_{0.75}$CoSb$_{0.85}$Sn$_{0.15}$ & I     & Ti$_{0.17}$Hf$_{0.79}$Co$_{1.01}$Sb$_{0.98}$Sn$_{0.06}$ & 9.849 \\
      & II    & Ti$_{0.60}$Hf$_{0.38}$Co$_{1.09}$Sb$_{0.27}$Sn$_{0.66}$ &\\
Ti$_{0.15}$Hf$_{0.85}$CoSb$_{0.85}$Sn$_{0.15}$ & I     & Ti$_{0.10}$Hf$_{0.86}$Co$_{1.03}$Sb$_{0.94}$Sn$_{0.07}$ & 9.705 \\
      & II    & Ti$_{0.47}$Hf$_{0.48}$Co$_{1.11}$Sb$_{0.29}$Sn$_{0.65}$ &\\
HfCoSb$_{0.85}$Sn$_{0.15}$    & I     & Hf$_{0.93}$Co$_{1.02}$Sb$_{0.97}$Sn$_{0.08}$ & 10.581 \\
      & II    & Hf$_{0.86}$Co$_{1.19}$Sb$_{0.26}$Sn$_{0.69}$ &\\
 \hline
\end{tabular}%
\end{table}

\begin{figure}[ht]
\centering
  \includegraphics[width=12cm]{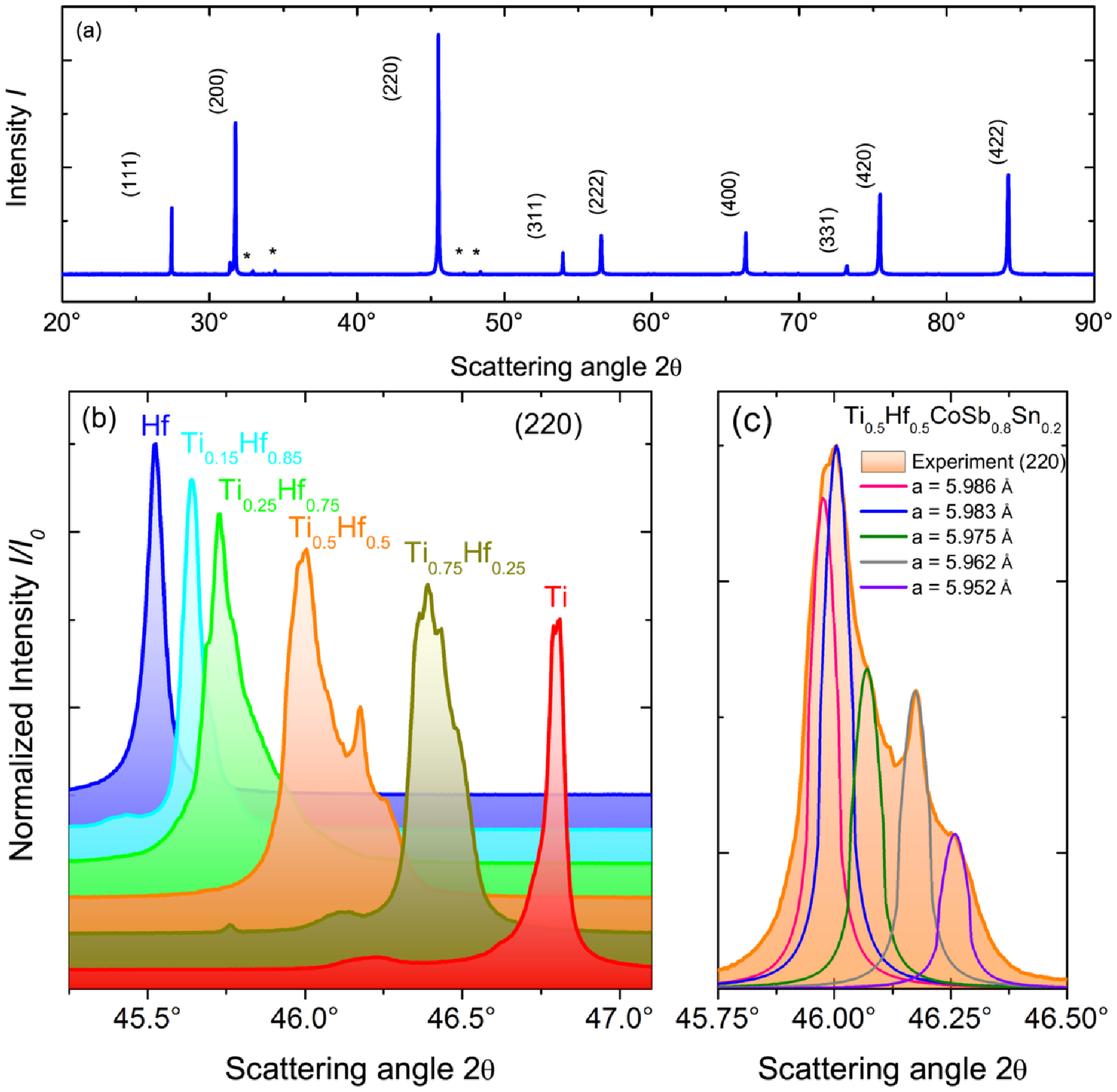}
  \caption{(a) X-ray powder diffraction pattern (XPD) of HfCoSb$_{0.85}$Sn$_{0.15}$ obtained using synchrotron 
  					radiation ($\lambda$ =1.65307~\AA{}; * indicates $\beta$-Sn). XPD of  the main reflection (220) 
  					of Ti$_{1-x}$Hf$_{x}$CoSb$_{0.85}$Sn$_{0.15}$ with the indicated ratios of Ti to Hf, (c) fitting of the 
  					(220) reflections for Ti$_{0.5}$Hf$_{0.5}$CoSb$_{0.85}$Sn$_{0.15}$.}
  \label{fgr:xrd}
\end{figure}

X-ray diffraction patterns of the substitution series Ti$_{1-x}$Hf$_{x}$CoSb$_{0.85}$Sn$_{0.15}$ 
using a Cu$K_{\alpha 1}$ lab source confirmed the presence of Half-Heusler structure C1$_{b}$ in all 
compounds despite the presence of a marginal amount of $\beta$-Sn close to the detection limit. Figure~\ref{fgr:xrd}(a) shows the representative XPD pattern of HfCoSb$_{0.85}$Sn$_{0.15}$ 
with the indexed reflections.
However, high-resolution XPD with synchrotron radiation revealed the existence of a multiphase state in samples with x $\neq$ 0,1, thus confirming phase separation into different Half-Heusler phases, as observed by EDX analysis. 
For EDX spectroscopy, several points on a limited image section were chosen. Thus, only two Half-Heusler phases were distinguishable for each samples.
Samples with a single element Ti ($x= 0$) or Hf ($x=1$) exhibited sharp and symmetric 
Bragg reflections, thereby indicating the presence of only one phase with the C1$_{b}$ structure. 
The reflections for a composition of $x = 0.15$ show very 
little broadening of the peaks. In contrast, 
the splitting of the main reflection (220) becomes very obvious for 
samples with 0.25 $\le$ x $\le$ 0.75 (see Figure~\ref{fgr:xrd}(b) due to the presence of several Half-Heusler phases. 
For example, the fitting of the powder pattern of Ti$_{0.5}$Hf$_{0.5}$CoSb$_{0.85}$Sn$_{0.15}$ 
requires at least the presence of five different Half-Heusler phases with very similar lattice 
parameters $a$ (see Figure \ref{fgr:xrd}(c). As an initial assumption, we used the composition of 
the Half-Heusler phases determined by EDX spectroscopy (see Table~\ref{tbl:edx}) and refined the lattice parameters. 
Subsequently, more phases with the C1$_{b}$ structure were added to model the shape of 
the (220) reflection correctly.

\subsection{Thermoelectric properties}
\begin{figure}[b]
\centering
   \includegraphics[width=8cm]{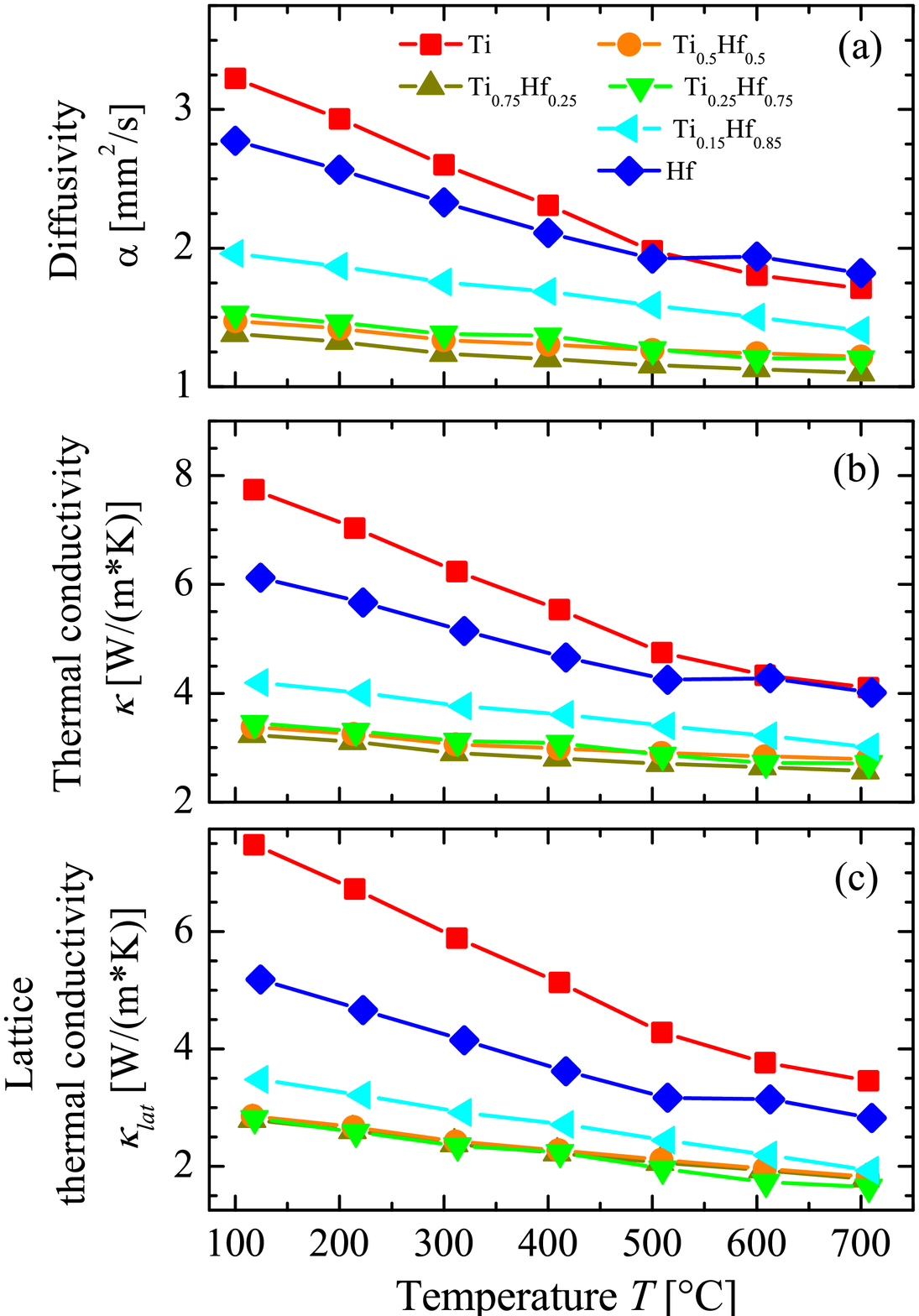}
  \caption{ (a) Diffusivity $\alpha$, (b) thermal conductivity $\kappa$, and (c) lattice thermal conductivity 
  					$\kappa_{lat}$ as functions of temperature for Ti$_{1-x}$Hf$_{x}$CoSb$_{0.8}$Sn$_{0.15}$ for the indicated ratios of Ti to Hf.}
  \label{fgr:kappa}
\end{figure}

The diffusivity and thermal conductivity plots of the investigated compounds 
are displayed in Figure~\ref{fgr:kappa}. 
These values are highest for single-phase samples with $x = 0$ and $x = 1$. 
The diffusivity and thermal conductivity are effectively suppressed by the 
intrinsic phase separation. This suppression is independent of the exact composition $x$. 
All samples with a Hf content of 0.25 $\le$ $x$ $\le$ 0.75 exhibit very 
similar thermal conductivities. The values are nearly temperature-independent, 
thereby indicating a glass-like behavior. 
A slightly higher thermal conductivity was observed for samples with $x = 0.85$. 

To further examine the impact of the substitution level, we calculated the lattice 
thermal conductivity by applying the Wiedemann--Franz law (see Figure~\ref{fgr:kappa}(c)). 
The lowest $\kappa_{lat}$ value of 1.64~Wm$^{-1}$K$^{-1}$ at 710$^{\circ}$C was reached by the sample with $x = 0.25$.
Samples with $x = 0.5$ and $x = 0.75$ exhibit values of 1.82~Wm$^{-1}$K$^{-1}$ and 1.78~Wm$^{-1}$K$^{-1}$, 
respectively, while the samples with Hf content of $x = 0.85$ exhibited the highest lattice 
thermal conductivity among all phase-separated samples (1.93~Wm$^{-1}$K$^{-1}$). 
The corresponding BSE images show that these results correspond with the observed microstructuring 
of the samples. The network of the second Heuser phase in the matrix is well-formed 
in Ti$_{0.25}$Hf$_{0.75}$CoSb$_{0.85}$Sn$_{0.15}$ (see Figure \ref{fgr:sem}(d)). 
Meanwhile in samples with $x = 0.5$ (see Figure~\ref{fgr:sem}(c)) and $x = 0.25$ (see Figure~\ref{fgr:sem}(b)) 
the amount of secondary phase reduces and the samples exhibit larger areas of the single-phase matrix. 
With a further increase of Hf ($x = 0.85$, see Figure~\ref{fgr:sem}(e)), the 
amount of the secondary Half-Heusler phase is so small, that it cannot form a interconnecting network through the matrix anymore. This agrees well with the results from the high-resolution XPD. The shape and width of the (220) reflection for $x = 0.85$ is similar to that of the single phase samples in contrast to the splitting of the main reflection for samples with 0.25 $\le$ $x$ $\le$ 0.75. Therefore, phonon scattering is most 
effective in the sample with $x = 0.75$ corresponding to a reduction of 52\% in 
the lattice thermal conductivity when compared with that of the single-phase TiCoSb$_{0.85}$Sn$_{0.15}$. 

\begin{figure}[h]
\centering
   \includegraphics[width=8cm]{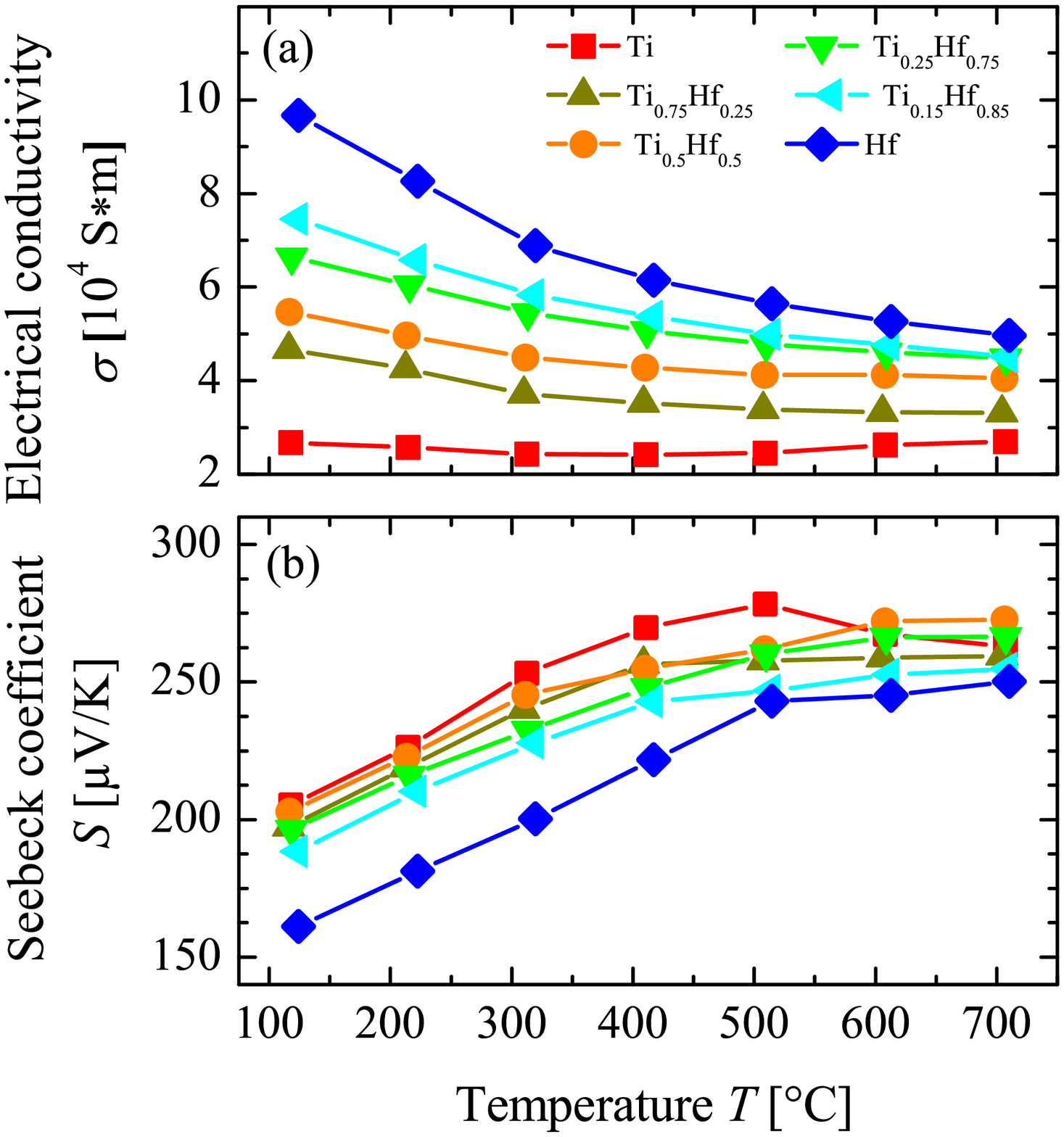}
  \caption{(a) Electrical conductivity $\sigma$ and (b) Seebeck coefficient $S$
  as functions of temperature for Ti$_{1-x}$Hf$_{x}$CoSb$_{0.85}$Sn$_{0.15}$
  with the indicated ratio of Ti to Hf.}
  \label{fgr:PF}
\end{figure}

Figure~\ref{fgr:PF}(a) shows the temperature dependency of the electrical 
conductivity $\sigma$, for all the samples. All samples exhibit a metallic behavior, i.e.,
the conductivity decreases with increasing temperature. Upon substitution of Ti with its 
heavier homologue Hf, $\sigma$ is enhanced.
Due to the interrelation of the Seebeck coefficient with $\sigma$, the 
Seebeck coefficients also shows a decrease upon substitution. 
The maximum Seebeck coefficient was observed for TiCoSb$_{0.85}$Sn$_{0.15}$, 
which value was 278~$\mu$V/K at 500$^{\circ}$C. 
This maximum is shifted to higher temperatures upon substitution of Ti by Hf. 
Even though the substitution is isoelectronic, it changes the electronic 
structure and consequently affects the  carrier concentration and effective mass of the charge carriers. 
Nevertheless, the Seebeck coefficients and electrical conductivity values of all substituted 
samples lie in between the values for the unsubstituted ones and vary as 
expected with the substitution level. 
Therefore, we conclude that the electronic properties are not affected by the 
phase separation.

The calculation of the power factor $S^2\sigma$ reveals that the optimal electronic 
properties are obtained for samples with a 
Hf content of 50\% and more. At 710$^{\circ}$C, we obtained power factors between 2.93~mWK$^{-2}$m$^{-1}$ ($x = 0.15$) 
and 3.18~mWK$^{-2}$m$^{-1}$ ($x = 0.25$).
These values correspond to a power factor enhancement of 89\% with respect to that of TiCoSb$_{0.85}$Sn$_{0.15}$ (1.68~mWK$^{-2}$m$^{-1}$ 
at 710$^{\circ}$C). 

\begin{figure}[h]
\centering
   \includegraphics[width=10cm]{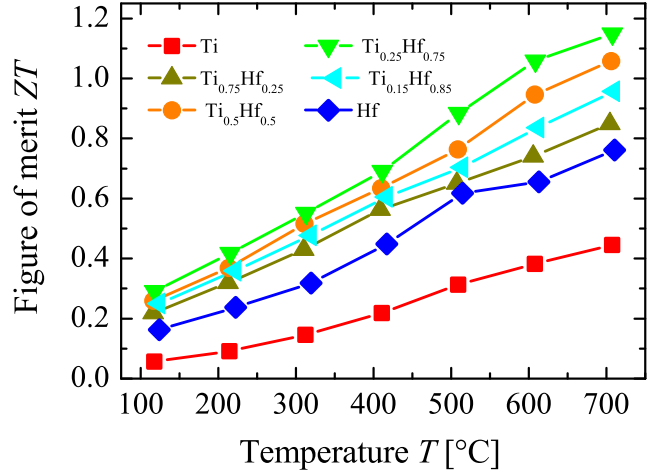}
  \caption{Figure of merit $ZT$ as a function of temperature for Ti$_{1-x}$Hf$_{x}$CoSb$_{0.8}$Sn$_{0.15}$ for the indicated ratios of Ti to Hf.}
  \label{fgr:ZT}
\end{figure}

From Figure~\ref{fgr:ZT} we note that fine-tuning of the Ti--Hf ratio has a significant 
impact on the figure of merit $ZT$. 
By simply mixing Ti and Hf together in a 50\% ratio, $ZT$ shows a immediate improvement by 42\% 
when compared with that of the sample with only Hf. 
Since both samples exhibit similar power factors, this improvement is due to reduction in the 
lattice thermal conductivity by intrinsic phase separation. Adjustment of the ratio of 
Ti to Hf enhances $ZT$ further. 
The selection of the optimal electronic properties in combination with the lowest thermal conductivity leads to 
the maximum figure of merit of $ZT = 1.15$ at 710$^{\circ}$C for Ti$_{0.25}$Hf$_{0.75}$CoSb$_{0.85}$Sn$_{0.15}$. 
This value clearly exceeds the benchmark of $ZT$ $\approx$ 1 for device applications.\citep{CR13} 
This result can be considered as very significant for bulk $p$-type Half-Heuslers. 
When compared with the previous state-of-the-art material, e.g., Ti$_{0.12}$Zr$_{0.44}$Hf$_{0.44}$CoSb$_{0.8}$Sn$_{0.2}$~\citep{YLC13} 
with $ZT = 0.98$ at 700$^{\circ}$C, this value corresponds to an improvement of 19\%.
Considering only ingot samples prepared by a simple arc melting fabrication 
process as in reference~\citep{CSP08}, 
the enhancement of the figure of merit exceeds 125\% by our combined approach 
of phase separation and carrier concentration adjustment.


\section{Conclusions}

Two concepts were successfully applied to improve the thermoelectric properties 
of the $p$-type TiCoSb system. Starting from Ti$_{0.5}$Hf$_{0.5}$CoSb$_{0.8}$Sn$_{0.2}$, we showed that phase separation ($ZT$ = 0.9 at 710$^{\circ}$C \citep{RBO14}) affords thermoelectric properties that are similar to 
those obtained with nanostructuring approach involving ball milling.\citep{YLW12} In the second step,
the optimization of the carrier concentration led to an improvement of 28\% as regards the 
figure of merit in Ti$_{0.5}$Hf$_{0.5}$CoSb$_{0.85}$Sn$_{0.15}$ ($ZT$ = 1.05). 
By subsequently adjusting the Ti to Hf ratio for optimum phonon scattering, we 
achieved a maximum $ZT$ of 1.15 at 710$^{\circ}$C in Ti$_{0.25}$Hf$_{0.75}$CoSb$_{0.85}$Sn$_{0.15}$. 
When compared with the $ZT$ value of the previous state-of-the-art $p$-type Half-Heusler material, 
Ti$_{0.12}$Zr$_{0.44}$Hf$_{0.44}$CoSb$_{0.8}$Sn$_{0.2}$ \citep{YLC13} with a ZT = 1.0 at 700$^{\circ}$C, 
this value corresponds to an improvement of 15\%.
In conclusion, we realized a bulk $p$-type Half-Heusler material with TE performance as good as 
the state-of-the art $n$-type Ti$_{0.5}$Zr$_{0.25}$Hf$_{0.25}$NiSn$_{0.998}$Sb$_{0.002}$ 
($ZT = 1.2$ at 560~$^{\circ}$C).~\citep{SB13}
As both materials perform well together in one TE module,~\citep{BBZ14} the 
application of TE devices based on phase-separated Half-Heusler 
is soon expected to be realized.


\section{Experimental details}

Ingots of ca. 15~g were prepared by arc melting of the stoichiometric amounts of the elements. 
For homogenization, each sample was crushed and remelted several times. At each step, the weight 
loss due to the high vapor pressure of Sb was compensated by addition of Sb until no evaporation 
was observed during the melting process. The samples were annealed in evacuated quartz tubes at 
900$^{\circ}$C for seven days and subsequently quenched in ice water. 
The crystal structure was analyzed by X-ray powder diffraction (XPD) with CuK$_{\alpha 1}$ 
radiation at room temperature using an image-plate Huber G670
Guinier camera equipped with a Ge(111) monochromator operating 
in the range 10$^{\circ}$ $\le$ 2$\theta$ $\le$ 100$^{\circ}$.
To investigate the phase separation in detail all samples were analyzed using high resolution 
X-ray powder diffraction with synchrotron radiation with $\lambda $ = 1.65307 \AA{}.
The experiments were performed at the XPD beamline at the bending 
magnet D10 at the Brazilian Synchrotron Light Laboratory (LNLS). For details regarding the 
characteristics of the beamline see, e.g., reference \citep{FGC06}.\\
Microstructures characterization were performed on metallographically
prepared cross-section of approximetaly 4$\times$1 mm$^{2}$. The phase distribution was documented by 
optical (Zeiss, Axioplan) and scanning electron microscopy (SEM, Philips XL30 with LaB6 cathode). 
The semi-quantitative determination of the chemical composition and the recording of the 
qualitative element distribution were realized by using the element specific X-ray intensities 
measured with an energy dispersive detector (Bruker, XFlash SDD 30~mm) that was attached to the 
SEM. An acceleration voltage of 25 kV and the spot mode were used for the determination of the 
chemical compositions of the phases. The PB-Phi(Rho-z) matrix correction model was applied 
for calculation of the composition (Bruker, Quantax 400 software package Ver. 1.9.4). 
Hypermaps with enhanced count rates were recorded overan interval 60 min, and the element-distribution 
images were extracted from the deconvoluted intensities of the X-ray lines. In case of Hf, 
the M$\alpha$ line instead of the L$\alpha$ line was selected to represent the Hf distribution 
due to the better spatial resolution obtained in the former case. For all other elements, the same X-ray lines were used for 
the quantification and the element distribution images.\\
To investigate the thermoelectric properties at high temperature, the ingots were cut into discs 
and bars. The Seebeck coefficients and electrical conductivity were determined simultaneously using an 
LSR-3 (Linseis). 
The thermal conductivity $\kappa$ was calculated using the relation $\kappa = C_p \alpha \rho$, 
where $C_p$ denotes the specific heat capacity, $\alpha$ the thermal diffusivity, 
and $\rho$ the density. The values $\alpha$ were measured by means of the laser flash method using 
the Netzsch LFA 457 instrument. The density $\rho$ was calculated from the mass and 
volume of the cut bars. The heat capacities were estimated by means of the Dulong--Petit law.
The uncertainties were 3\% for the electrical conductivity and 
thermal diffusivity and 5\% for the Seebeck coefficient, thereby leading to an 11\% 
uncertainty in the $ZT$ values.
We repeated the experiments numerous times and confirmed that the peak $ZT$ 
values were reproducible within 5\%.
Furthermore, we measured the samples while they were heated up to 700$^{\circ}$C and cooled, 
and we subsequently verified that there was no degradation in the various sample properties.\\

\section{Acknowledgements}

The authors gratefully acknowledge financial support by the DFG Priority Program 
1386 "Nanostructured Thermoelectric Materials under proposal BA 4171/2-2 and the thermoHEUSLER Project (Project No. 0327876D) of the German Federal Ministry of Economics and Technology (BMWi).
We thank Sylvia Kostmann for specimen preparation for the microstructural examination, and Monika Eckert and Petra Scheppan for SEM images and EDX measurements. 
We also thank the staff of the LNLS (Campinas) for support as well
as Dean H. Barrett (LNLS, Campinas) for help with the XRD experiments. 
This work was further supported by the Brazilian Synchrotron Light Laboratory
(LNLS) under proposals XPD - 17015 and by the DAAD (Project No. 57060637).

\bibliography{paper}

\end{document}